# Fugacity of condensed media

Leonid Makarov and Peter Komarov
*St. Petersburg State University of Telecommunications,
Bolshevikov 22, 193382, St. Petersburg, Russia*

**ABSTRACT**

Appearance of physical properties of objects is a basis for their detection in a media. Fugacity is a physical property of objects. Definition of estimations fugacity for different objects can be executed on model which principle of construction is considered in this paper. The model developed by us well-formed with classical definition fugacity and open up possibilities for calculation of physical parameter «fugacity» for elementary and compound substances.

## 1. INTRODUCTION

The reality of world around is constructed of a plenty of the objects possessing physical properties. Duration of objects existence – their life time, is defined by individual physical and chemical parameters, and also states of a medium. In the most general case any object, under certain states, can be in one of three aggregative states - solid, liquid or gaseous. Conversion of object substance from one state in another refers to as phase change. Evaporation and sublimation of substance are examples of phase changes.

Studying of aggregative states of substance can be carried out from a position of physics or on the basis of mathematical models. The physics of the condensed mediums has many practical problems which decision helps to study complex processes of formation and destruction of medium objects. Natural process of object destruction can be characterized in parameter volatility – fugacity. For the first time fugacity as the settlement size used for an estimation of real gas properties, by means of the thermodynamic parities deduced for ideal gases, was offered by Lewis in 1901.

Expanding these representations it is possible to specify, that all objects of a medium are material, so, for them there is a physical description. Following the general representations about fugacity as settlement physical parameter of the substance forming the condensed phase or entering into its structure, it is accepted to believe, that value of this parameter as much as possible in sated to prime steam of this phase. Such concept allows approving, that fugacity is the physical parameter of substance describing ability of substance of concrete object to conversion in a gaseous state, at constant parameters of medium. The most important parameters of medium are pressure and temperature. Generally fugacity it is possible to identify with a physical parameter of material of the research object, describing ability of substance to an output from the condensed phase as which the solid or a liquid is considered. It is possible to note, that in the physical understanding, the given term well corresponds to true. The Latin name of physical process of substance weight loss, entered Lewis, corresponds: "*fugitivus*" - escaped, departed, well explains an essence of the phenomenon. It's just as process of substance transformation in to the many particles state. We shall designate this process by the term «*fugacity*» and to denote of symbol «*F*».

Natural character of process fugacity as physical process is caused by thermodynamic axioms. So considering any object as thermodynamic system, it is possible to tell, that entropy systems ($S$) are characterized with quantity of possible microstates adjusted with their thermodynamic properties. (Ludwig Boltzmann. 1877).

$$S = k \ln \Omega \qquad (1)$$

where a constant $k=1.38 \cdot 10^{-23}$ *J/K*, and $\Omega$ is number of microstates which are possible in an available macroscopic state. This postulate, describes thermodynamic systems, using statistical representation about behavior of components making them. Besides, the postulate approves about presence of indissoluble communication between microscopic properties of object with one of thermodynamic properties. On this basis the conclusion is done that entropy is function of a state and characterizes a measure of the structural disorder.

Basis of such representation about investigated object makes entropy of transition. Assuming as a basis classical model of gases where the structural order is destroyed, it is possible to consider a state of thermodynamic system on boundary of phases: solid – gaseous or a liquid – gaseous as a state at extreme to a heat.

It is obvious, that to process in which the border of phase change is considered, entropy corresponds to value of boiling-point, defined individually for each substance. Statistical definition entropy is a basis for the formulation of the thermodynamic second law, which consequence is the statement that the thermodynamic system aspires to pass spontaneously in a state with the maximal thermodynamic probability.

The condensed mediums are widespread in the Nature. Typical examples of the condensed are liquids, crystals, amorphous substances and gaseous. The condensed medium – as object of research, can be represented by model which allows study some property of natural object. Within the limits of model (**Fig. 1**) it is possible to consider a solid state of object which can be connected with other aggregative states.

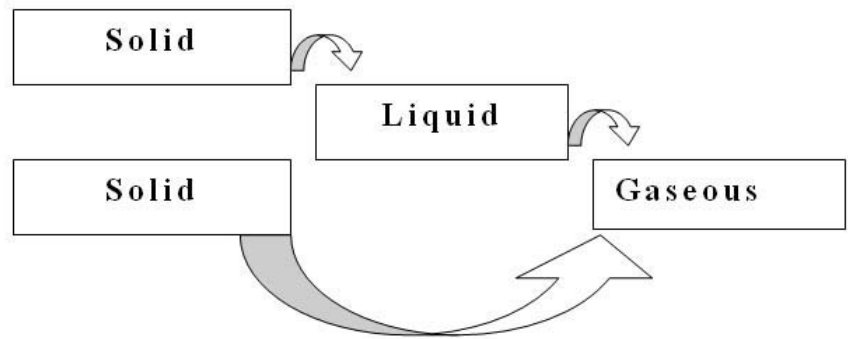

*Fig. 1. The scheme of model of aggregative states of substance*

The phenomenon of fugacity – is natural physical process, which basis makes emission of particles of object weight in surrounding space. For a liquids is an evaporation and for solids – sublimation.

## 2. THERMODYNAMIC SYSTEM MODEL "SUBJECT – MEDIUM"

From the position of molecular physics studying physical properties and aggregative states of various objects in view of their molecular structure is of interest. In theoretical researches of molecular physics two approaches are used. First of them is realized within the limits of statistical, and the second thermodynamic physics. For the statistical approach use of mathematical probability theory in studying macroscopic properties of objects is characteristic. The basis of the thermodynamic approach is made with the analysis of quantitative parities of the major parameters of object state.

The thermodynamic method, unlike a statistical method, is not connected with any concrete representations about an internal structure of objects and elements. The thermodynamics operates with macroscopic characteristics of studied objects. Assuming as a basis a little bit empirically established positions - laws of thermodynamics which possess rather big generality, we shall generate calculation fugacity principles. Representation about thermodynamic system is created on the basis of the general representations about the macroscopic elements considered in the form of uniform system. At studying evolution of thermodynamic system the big attention is given questions of a power exchange between elements of system and a medium. A typical example of thermodynamic system – is a liquid and adjoining it of pairs.

Let's define thermodynamic system ($A_i$) as a unique composition of the macroscopic elements, exchanging energy among themselves and a medium ($W$). We shall consider model (***Fig. 2***) thermodynamic system $A_i$, which can be in different aggregative states ($e_1$, $e_2$). We believe, that the system is characterized by a set of macroscopic elements, which exchange energy both inside of system, and with an medium, in which essential parameters are presented by pressure and temperature ($P$, $T$). During supervision over succession of events in the medium there is an opportunity to find out a particle of weight ($\Delta m$), belonging $A_i$ which are any way injected in surrounding space.

Considering the power characteristic of system, we shall specify on two components: a parameter of power communication between elements of system and a power parameter of external processes. Despite of presence of the certain analogy in definition of the specified parameters here it is easy to find out essential distinctions. In the first case the generalized parameter of power communication which characterizes of a system phase state is allocated. In the second case the power parameter of system conversion process from a solid or liquid aggregative state in steam – a gaseous aggregative state is considered.

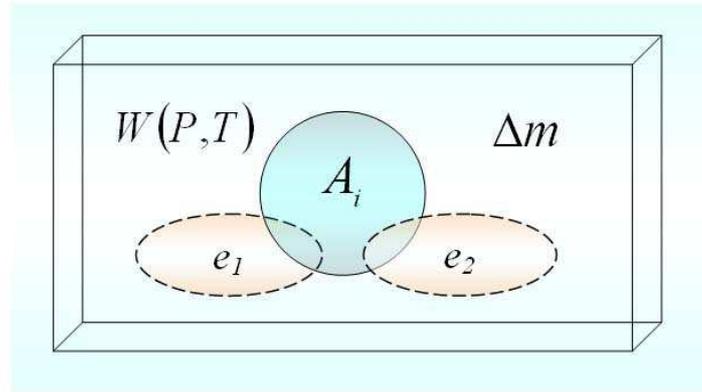

*Fig. 2. Thermodynamic system model "Subject – Medium"*

In the most general case it is possible to tell, that the state of thermodynamic system is defined by set of values of thermodynamic parameters: density, polarization, magnetization, boiling, freezing and melting points. We shall enter a designation: boiling point as $T_{BP}$, melting point as $T_{MP}$ and freezing point as $T_{FP}$. It is possible to expand the list of these parameters. It is obvious, that the majority of the specified parameters are characterized with macroscopic properties of system. From here follows that two states of system admit different if for them numerical values even one of thermodynamic parameters are various.

If the system state does not change in time then it is permanently. The stationary state of system refers to equilibrium if its invariance in time is not caused by action of any, external process. We believe that thermodynamic parameters of system are interconnected. Therefore the equilibrium state of system can be defined unequivocally, by means of the task of values of the limited number of these parameters. In thermodynamics distinguish external and internal parameters of a system state. External parameters of a state the parameters depending only from generalized coordinates of external bodies or factors with which the system cooperates refer to. We shall allocate parameter of external temperature – ambient temperature $T_{AT}$. Internal parameters of system we shall set values of boiling and melting points. The specified parameters allow identifying various objects of a medium generally.

## 3. THE NECESSARY STATES OF BEING THERMODYNAMIC SYSTEM MODEL "SUBJECT – MEDIUM"

At studying physical properties of object (substance) the big attention is given definition of parameters of a state. In particular, allocate the parameters describing change of aggregative states: gaseous, liquid or solid. As is known, if to choose system of coordinates: pressure ($P$) – temperature ($T$) it is possible to display curves of balance between various phases of the given substance. The illustrative image of the phase plot is shown on (***Fig. 3***).

The trajectory $OT_{TP}$ corresponds to balance between a solid and gaseous phase, trajectory $T_{TP}B$ corresponds to balance between a liquid and the steam, and trajectory $T_{TP}M$ corresponds to balance between a solid and liquid phase. Such representation about behavior of thermodynamic system appears useful to draw a conclusion on existence of two important state parameters: melting point ($T_{MP}$) and boiling point ($T_{BP}$).

Really given parameters define a power state of thermodynamic system which can be considered as the identifier of object. In this case for each object (substance), at constant pressure ($P$), it is possible to specify pair values ($T_{MP}, T_{BP}$) for which there is a unique object (substance).

It is possible to allocate the basic fragments. Line $OT_{TP}$ corresponding balance between solid and gaseous phases, refers to sublimation curve. Line $T_{TP}B$ of balance between a liquid and the steam refers to as a curve of evaporation. Line $T_{TP}M$ of balance between a solid and a liquid refers to as a curve of fusion.

The specified parameters of object state allow allocating phase's boundary (***Fig. 4***).

Let's note the phase states responding to fugacity process and corresponding declarative temperature parameters: a solid – gaseous ($\{T_{MP}, T_{BP}\} \to \{T_1, T_2\}$); a liquid – gaseous ($\{T_{MP}, T_{BP}\} \to \{T_1, T_3\}$).

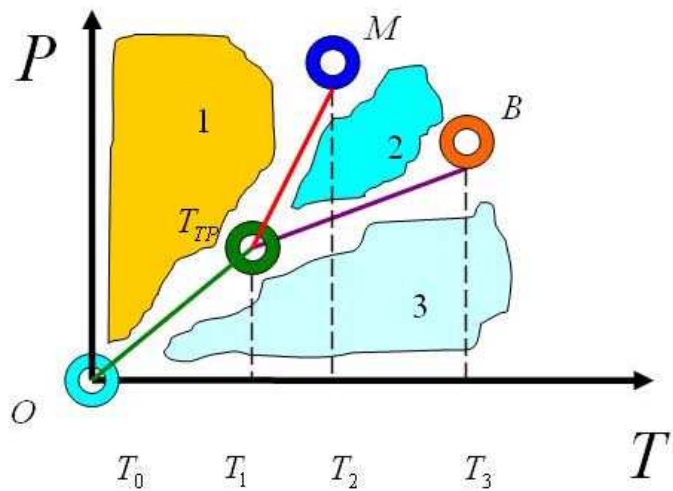

***Fig. 3.*** *Typical substance phase plot*
*$T_{TP}$ – triple point. Area 1 – solid phase,*
*area 2 – liquid phase, area 3 – gaseous phase.*

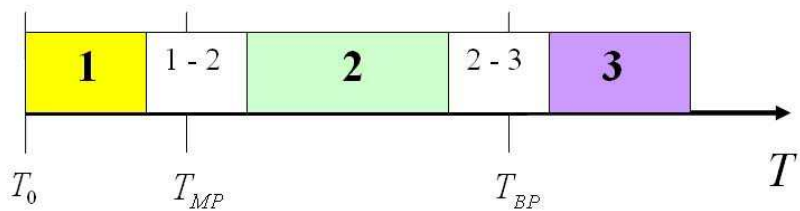

***Fig. 4.*** *Scheme of aggregative states.*
*Boundary: 1-2 area a solid – liquid;*
*2-3 liquid – gaseous*

Let's formally allocate a rule according to which for each physical object $A_i \in L$ always it is possible to specify pair temperature values $\{T_{MP}, T_{BP}\}$ which irrespective of an aggregative state allow leading identification of object. A basis for creation of such rule Pauli-Fermi principle which approves impossibility of existence in atom of two electrons with identical quantum numbers can be. Taking it into consideration it is possible to tell, that pair values - $\{T_{MP}, T_{BP}\}$ creates the power image of object generated on the basis of quantum representations about structure of substance. The expanded representations about communication of topology with power parameters of object are established in physical chemistry – stereochemistry. However, the given rule is necessary, but not sufficient to carry out procedure of full identification. Validity of this remark we shall disassemble on a concrete example.

For some substances the parameter pressure $P$ in a triple point is less then atmosphere pressure *1atm=101325 Pa*. Normal atmospheric pressure name pressure *P=101325 Pa*. Such substances at heating in normal atmospheric pressure are melted. For example, the triple point ($T_{TP}$) of water has coordinates $T_{TP} = 273.15K$

This point is used as basic for calibration of an absolute temperature of Kelvin scale. Exist, however, and such substances at which triple point pressure $P_{TP}$ exceeds $P = 1atm$. So for carbonic acid $CO_2$ it is possible to specify typical physical parameters: pressure $P_{TP} = 5.11 atm$ and temperature $T_{TP} = 216.55K$. As it is easy to notice, these parameters differ from physical parameters for the typical (normal) medium at which are spent a plenty of supervision over objects of medium. Therefore at normal atmospheric pressure carbonic acid snow can exist only at low temperature, and in a liquid state at $P = 1atm$ this object does not exist. In balance with their vapor at atmospheric pressure ($P = 1atm$) carbonic acid is at temperature 194.65K or –78.5°C in a solid state. It is widely applied «artificial ice», which never melted but only sublimated.

Taking it into consideration, definition fugacity we shall spend under normal states an medium: $P = 1atm$, $T_{TP} = 273.15K$. If it is required to define a parameter fugacity in other states, in particular, at other ambient temperature such procedure will be considered in addition. This explanatory is necessary because there are distinctions in definition of standard states of solid and liquid, on the one hand, and gaseous, on the other hand. The standard state for a gaseous element or connection is defined not how for solid substance or a liquid. For a standard state of ideal gaseous this satisfies to parity:

$$PV = nRT \tag{2}$$

accept considered gaseous at the given temperature and pressure $P = 1atm$, which behavior corresponds to ideal gas.

## 4. PARAMETERS OF PHASE CHANGES

Let's consider phase changes. The crystal substance is characterized by the certain natural arrangement of particles making it. For a liquid and gaseous infringement of the order of an arrangement of elements of structure is characteristic. In standard condition, at $P = 1atm$ and ambient temperature $T_{AT} = 273.15K$, the physical object loses a part of the weight. This is demonstration of fugacity process. At the established ambient temperature and furthermore at consecutive heating solid substance energy of oscillatory movement of its particles all time increases therefore their mutual pushing apart amplifies also. Some particles of substance have an opportunity to break forces of an attraction and pass in surrounding space. This is phenomenon fugacity "a solid – a medium".

For each substance which is being a solid phase, there is a certain melting point $T_{MP}$ at which the attraction of particles to each other weakens. The substance is melted also passes in a liquid state. In a liquid the mutual attraction of molecules still is enough to keep them together, and only separate, is the fastest at present to moving molecules it is possible to come off a surface. This is phenomenon fugacity "a liquid – a medium".

At the further heating a liquid the number of fast molecules increases. Pressure pair the given substance above a surface of a liquid gradually increases. For each liquid there is an individual value of temperature of boiling $T_{BP}$ at which pressure the pair becomes equal to external pressure. The liquid begins to boil. Above point $T_{BP}$ exists only steam – a gaseous phase.

Allocating the phase change describing formation of gaseous fraction of investigated object, that postulates change of the condensed phase which is reflected in loss of some part of weight of object – to emission of particles. Following these representations it is easy to understand physical sense of a parameter летучести – fugacity. From this point of view fugacity is a parameter describing ability of substance, at the established values of pressure and temperatures of an medium, to lose a part of the weight. The parameter fugacity can be considered as an estimation describing ability of substance to inject a part of the weight in surrounding space. The disintegration of the condensed phase identified with one three phase states, occurs according to the law of increase entropy.

Transition in a new state of system is accompanied by change of internal energy. Internal energy of system (object) U - is defined as full energy minus kinetic energy of object, as the whole, and, potential energy of object in an external floor of forces. Hence, internal energy develops of kinetic energy of chaotic movement of molecules, potential energy of interaction between them and intramolecular energy.

Internal energy U is unequivocal function of a state of system which is characterized enthalpy $H$. It means that every time when the system appears in the given state, internal energy accepts value inherent in thus state, irrespective of background of system. Hence, change of internal energy at transition from one state in another will be always equal to a difference of values these states, irrespective of a way on which transition was made. Then the power state of system can be set concrete value:

$$\Delta H = H_2 - H_1 \tag{3}$$

On this basis it will be fair to specify an opportunity of synthesis of new objects which can be spent with use of the different technologies providing reception of identical objects on properties.

Considering, that it is impossible to measure internal energy of object directly, and it is possible to define only change of internal energy $\Delta U$ we shall write down:

$$\Delta U = Q - W \tag{4}$$

where: Q – internal energy bodies (heat) [J], W – work which should be made for translation of system in an medium [J].

It is possible to note, that destruction of system as working process, is identical to work on translation of elements of system in a medium. Fugacity – the physical parameter of thermodynamic system describing work which is made above system at transition from a solid or liquid state in a gaseous state, at constant external pressure. Such work made above system on translation from one aggregative state in another, is characterized by the energy identified with a physical parameter fugacity. Considering, that values of temperatures of phase changes for each substance are individual, the value of estimation received as a result of calculation fugacity also individually.

Formally, such conclusion can be received on the basis of calculation of change entropy thermodynamic system at heating from absolute zero up to temperature $T = T_{AT}$ at constant pressure $P$. As a basis we shall put:

$$\Delta Q_p = C_p dT \tag{5}$$

$$dS = \frac{\Delta Q_p}{T} \tag{6}$$

where: $\Delta Q_p$ - an elementary increment of internal energy of system; $C_P$ - isobaric a thermal capacity of system.

From here it is possible to establish, that

$$dS = C_p \frac{dT}{T} = C_p d(\ln T) \tag{7}$$

Considering, that $S_{T=0} = 0$

Let's receive expression for calculation entropy on the selected interval of temperatures:

$$S_T = \int_0^T C_P d(\ln T) \tag{8}$$

We believe that at $T = T_0 = 0K$ any substance is only in a solid state. In that case at heating substance transition in liquid or in a gaseous state is possible. For the phase changes occurring in isobaric-isothermal states, change entropy to equally resulted heat of phase change:

$$\Delta S_{PC} = \frac{\Delta H_{PC}}{Ò_{PC}} \tag{9}$$

where $H_{PC}$ - phase change enthalpy; $T_{PC}$ temperature of phase change ($Ò_{MP}, Ò_{BP}$)

On the basis of these reasoning it is possible to draw a conclusion, that heating of substance without phase changes is accompanied by continuous growth entropy. The description of such process is characterized by smooth function. At achievement of the temperature corresponding one phase changes there is a jump in change entropy. The physical essence of such process is obvious and can be explained by expression for calculation absolute entropy substances at any ambient temperature $T_{AT}$:

$$S_T = \int_0^{T_{MP}} \tilde{N}_{P,Sol} d(\ln T) + \frac{\Delta H_{MP}}{\grave{O}_{MP}} + \int_{\grave{O}_{BP}}^{\grave{O}_{BP}} C_{D,Liq} d(\ln T) + \frac{\Delta H_{BP}}{\grave{O}_{BP}} + \int_{\grave{O}_{BP}}^{\grave{O}} C_{P,Gas} d(\ln T) \quad (10)$$

Graphic dependence of change entropy $S_T$ thermodynamic system from temperature $T$ we shall present on *Fig. 5*. The received result allows drawing a conclusion, that absolute value entropy is defined by exclusively concrete values $T_{MP}, T_{BP}$ and mutually connected values $\Delta H_{MP}, \Delta H_{BP}$.

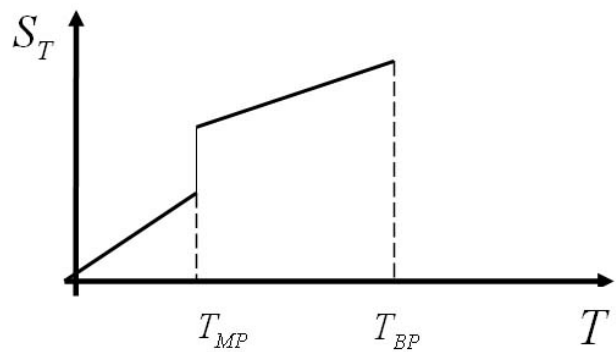

*Fig. 5. Modification of phase change entropy*

### 5. CONSTRUCTION OF FUGACITY'S PROCESS MODEL

Designing initial model of process, the postulate was established: action of the external factor – temperatures of medium, changes a state of object. Consequence of such statement is that at rise in ambient temperature, the object which is being the medium, changes by means of loss of weight. This phenomenon is certain as fugacity. At achievement of some critical temperature, characteristic only for the given object, there is a transition in a new aggregative state.

The formalism of model is well illustrated by expression (10). Here the effect of gradual destruction of investigated object, to be exact speaking its structural properties is distinctly observed. Believing, that such destruction is connected with gradual transition in a new aggregative state, distinctly we establish presence of smooth transition. It is possible to note, that action of the external factor – temperatures of medium, it is interfaced to the work directed on infringement макро of communications of elements of structure of object. Allocating this aspect it is possible to note, that accentuation of attention on the temperature factor of a medium does not settle all variety of possible revolting factors of medium. In the most general case here can be considered and other factors. However within the limits of the established model the temperature factor of medium which defines process fugacity is considered only. Influence of ambient temperature on process fugacity is considered at constant pressure which is accepted equal to atmospheric pressure. Introduction of such state allows receiving a set of comparable estimations fugacity, that creates a basis for construction phenomenological of some objects as which chemical elements can be considered, inert gaseous and various substances on a chemical compound, being in the certain aggregative states.

Following these representations, we shall consider the basic arguments creating a basis of designing of model of process fugacity, and also estimations of display of this process providing calculation in the certain states of a medium. We shall allocate the main arguments and we shall list them in such sequence.
1. Really in an expert it is easy to find out a plenty of substances with equal molecular weights, but different stereochemical structure. For such substances it is easy to find out distinctions in values of temperatures of fusion and boiling. Distinction of values of these parameters convincingly testifies to a different stereochemical structure of substances, despite of presence of the general chemical formula of their representation. If to use concept of a sign, as some identification symbol reflecting structural properties of substance values of temperatures of fusion and boiling can be a basis for construction of an image of substance (object).
2. Following the general physical representations it is possible to tell; that at absolute zero should the "rigid" structure of substance is observed. As opposed to it on a scale of temperatures it is possible to find such temperature point at which any substance exists in the form of plasma. It means, that fugacity as some function of model, is realized on the limited range of definition where limiting values of physical process fugacity can be established. Taking it into consideration it is possible to tell, that an image of substance as the formal model, has the limited area of representation. Then it is necessary to recognize presence of the bottom and top border of existence of an image that will well be coordinated with initial representations of physics. In such understanding the estimation fugacity can be formed on the basis of property of stability (stability) of substance – abilities to not collapse in the certain range of values of temperatures of a medium. A priori the top values of ambient temperature are not limited. Hence, it is possible to postulate equality of estimations fugacity for all substances, at great values of a parameter $T_{AT}$.
3. Really, the parameter of stability is connected with entropyей phase changes of substance under the selected external states. If as states of existence of object to consider only temperature of medium and pressure that will well be coordinated with the physical and chemical laws describing realization of different processes it is possible one of parameters to accept for a constant. We shall put as a constant pressure is accepted. Then stability – stability of substance destruction (change of an aggregative state), is characterized by values of temperatures of fusion and boiling. Within the limits of considered model it also will well be coordinated with initial representation about development of numerous processes of transition of substance from one aggregative state in another.

Believing, as before, that values of parameters $T_{MP}, T_{BP}$ are unique parameters for each substance $A_i \in L$, we have:

$$\Delta S_{A_i} \Rightarrow (T^{A_i}{}_{MP}, T^{A_i}{}_{BP}) \qquad (11)$$

where $\Delta S_{A_i}$ characterizes change entropy on an interval $T_{BP} - T_{MP}$.

Thus, we receive, that estimation entropy, calculated on an interval of temperatures in which the pair values $T_{MP}, T_{BP}$ gets, can be a basis for calculation of estimations fugacity. On the one hand the given conclusion will well be coordinated with the general physical representations about uniqueness of spatial structure of substance and presence of communication with power parameters of system. On the other hand, communication of power parameters of system with the work made at transition of system from one aggregative state in another, in particular, is found out is possible to allocate work on injection of weight of substance in surrounding space. Generalizing this thesis it is possible to tell, that the considered estimation entropy full enough and convincingly testifies to an opportunity of calculation of estimations fugacity for different substances on the basis of typical physical parameters of a state of objects of a medium.

### 6. INFORMATION MODEL OF SYSTEM "SUBJECT – MEDIUM"

Agreeing with a physical reality of existence of objects and recognizing uniqueness of values of parameters: $T_{MP}, T_{BP}$, we shall note their indissoluble communication with topology of an atomic design of substance, in other words, with stereochemical image of substance. This state allocates a unique set of temperature parameters. As it was specified earlier, this state is necessary. For completeness of object identification – substances the knowledge of molecular weight - $M$. Combination of a set of values $T_{MP}, T_{BP}$ and $M$ is required, allows to identify objects – substances. This state is full and sufficient. Following these representations, we shall write down expression for an estimation $St$ of stability of substance in the form of:

$$St = \frac{\chi[(T_{BP} - T_{AT}) - T_{MP}]}{M} = \frac{\chi A}{M}, [\frac{erg}{g/mol}], \quad (12)$$

where $St$ an estimation of stability (stability destruction at the established temperature of an medium);

$\chi = 1*10^{15}$ - coefficient;

$T_{AT}$ - ambient temperature;

$M$ - the molecular weight of substance expressed in gram/mol, numerically equal to relative molecular weight – gram.

The estimation of stability characterizes стационарность the object capable on the established time interval to keep constant internal energy.

Let's address to consideration of the information model which are clearing up main principles of calculation of an estimation fugacity.

Choosing some object $A_i \in L$, with the purpose of an establishment of an estimation fugacity, as the aprioristic information on object its physical parameters are used. Formally such parameters can be considered in the form of a sign design of the statement constructed in terms of physical parameters, for example, of temperature and weight. Believing, that the selected parameters allow receiving knowledge of individual property of object, it is required to establish the form of a sign design which provided carrying out of demanded calculations. In our case it is possible to consider two ways of reception of a required estimation fugacity. In one case that corresponds to the first way, the stationary medium, with the established typical parameters is considered.

The estimation fugacity pays off under constant states of medium for the selected substance possessing unique physical values of typical parameters of a state. In other case that corresponds to the second way, process of calculation of an estimation fugacity is realized similarly, but at change of ambient temperature. The choice of change of values of a temperature range of a medium is spent in view of properties of real objects. Not breaking the general line of reasoning, we can note, that expected downturn of values of a required estimation $St$ at increase in temperature of medium, should have the given reason limit. For this reason, despite of the big range of possible temperature values $T_{AT}$, for concrete substance, has no physical sense to consider values $St$ at $T_{AT} > T_{MP}$ as at $T_{AT} \gg T_{MP}$ substance of object passes in gaseous – plasma. These recommendations will allow reducing volume of settlement estimations if it does not contradict interests of research of the selected object. Strengthening this representation it is possible to tell, that at great values of temperature of medium all the substances placed on this medium, will have practically identical fugacity. These representations can be reflected in the form of information model (*Fig. 6*).

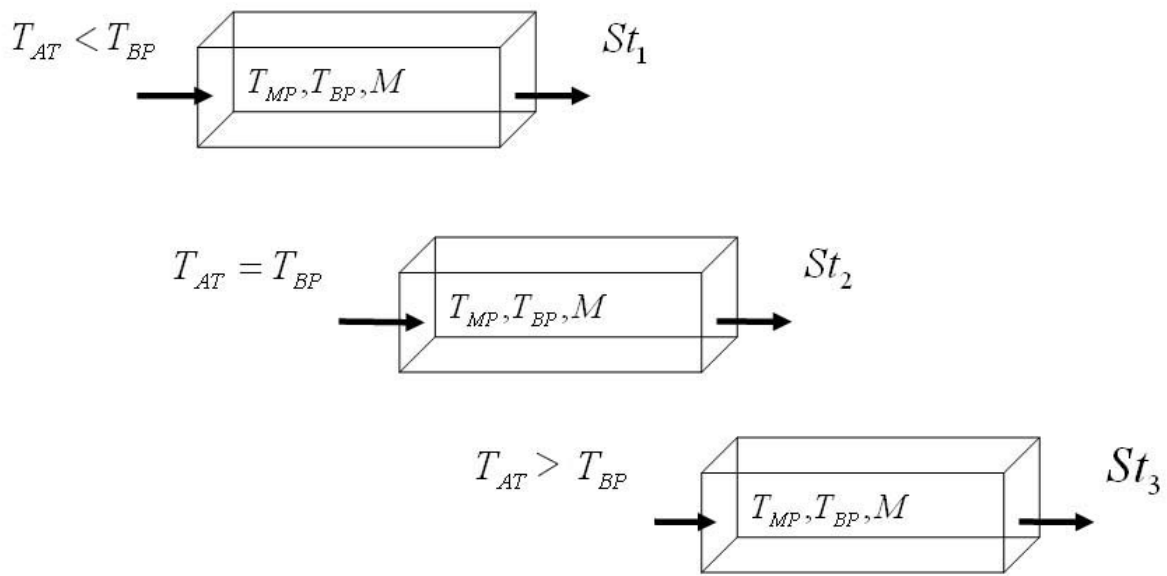

*Fig. 6. The information model scheme
of system "Object – Medium" states changes*

Obviously, with rise in temperature of medium intensity of process of destruction of investigated object increases, and, hence, increases fugacity. Within the limits of model, loss of substance of object, at change of temperature of medium, it is possible to consider on an interval of temperatures $[T_0; T_\infty)$. We believe, that at $T_0$ all objects are presented in a solid aggregative state, and temperature values $T_\infty$ are not limited. Then, for any objects of research it is possible to specify value of temperature $T_{AT}$ of a medium - $T_0 < T_{AT} < T_\infty$ at which estimations fugacity various objects are close or equal. It is obvious, that such limiting case corresponds to a unique state of substance – to plasma. In such states of a medium the state is executed, $T_{BP} << T_{AT}$ and investigated objects have the maximal estimation fugacity, and, hence, the maximal losses of weight of substance.

## 7. ELEMENT'S FUGACITY VALUATION

Research of physical properties of various objects has a good theoretical basis. So, for example, at research of properties of materials диэлектриков as substances with various physical parameters, the method of calculation of dielectric losses is widely applied. The given method of calculation of an estimation of dielectric losses possesses the certain analogy to calculation fugacity. The estimation of dielectric losses, with reference to various materials – to substances, in wide understanding of this term, is created on illustrative model in configuration space. We shall enter into consideration configuration space with system of coordinates: «Temperature» and «weight». Each object, for which the estimation fugacity is calculated, can be presented in the selected system of coordinates by three parameters: $T_{MP}, T_{BP}, M$. Model of succession of events in system: the object – an medium is formed on the basis of changes of temperature of medium. We shall lead positioning some object, having established values of physical parameters on corresponding axes of coordinates (*Fig. 7-a*). We shall lead graphic constructions (*Fig. 7-b*). We shall note a triangle $_\Delta OMT$ and an angle $\angle \beta$. We shall enter into consideration an angular measure. It is obvious, that

$$arctg(St) = \beta \tag{13}$$

Then an estimation fugacity we shall define in the form of:

$$F = \alpha - arctg(St) = \alpha - \beta \tag{14}$$

where $\alpha = 90^\circ = const$,

$F$ - is physical equivalent fugacity which is certain on an interval in radian measure $[0; \pi]$ or in grade measure $[0; 180]$.

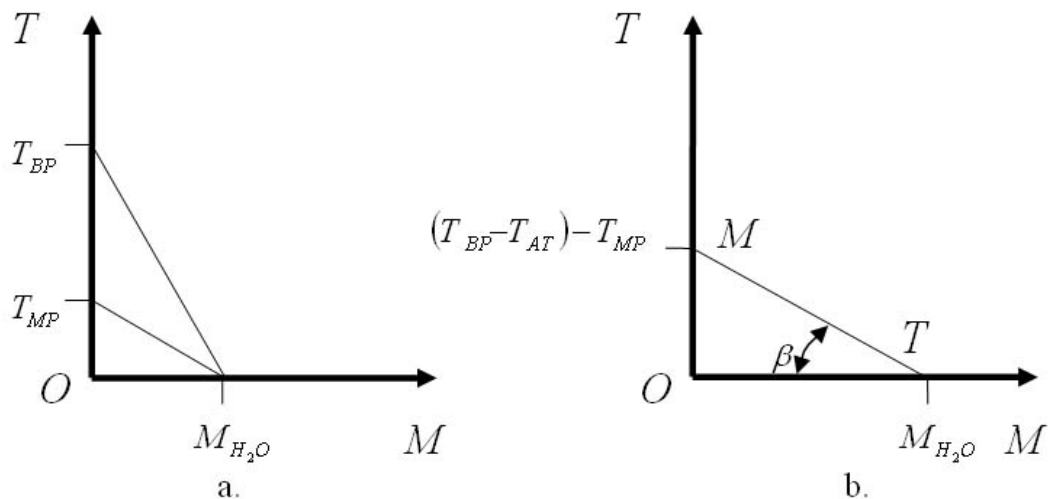

***Fig. 7.** Object positioning scheme*

On the basis of the considered model of fugacity definition as the settlement physical parameter describing ability of substance to collapse at different values of ambient temperature, we shall note:

1. Fugacity – property of all material objects possessing an opportunity of existence in two aggregative states;
2. Objects – substances for which the parameter fugacity pays off, possess a unique stereochemical design of association of chemical elements;
3. Objects – material objects are characterized by value of molecular weight and for them thermodynamic values of temperature of phase changes are known.

To interested of define estimation fugacity chemical elements of periodic system. With this purpose calculation of an estimation of stability we shall lead on expression (12), and an estimation fugacity on expression (14).

Let's address to chemical elements of periodic system. We shall display chemical elements according to their arrangement in periodic system (*Table 1*), we shall lead necessary calculations ($T_{AT} = 273.15K$) and we shall present a number of estimations fugacity.

*Table 1. Chemical elements.*
*Values of estimation St and F are given at $T_{AT} = 273.15K$*

| № | Element | Mass [g/mol] | $T_{MP}$ [K] | $T_{BP}$ [K] | St [ferg/g] | F [rad] |
|---|---|---|---|---|---|---|
| 1. | H | 1 | 17,81 | 20,28 | 0,512 | 1,098 |
| 2. | He | 4,002 | 0,95 | 4,216 | 0,169 | 1,403 |
| 3. | Li | 6,941 | 453,69 | 1620 | 34,805 | 0,029 |
| 4. | Be | 9,0122 | 1560,15 | 2745,15 | 27,236 | 0,037 |
| 5. | B | 10,811 | 2352,15 | 4273,15 | 36,806 | 0,027 |
| 6. | C | 12,011 | 4098,15 | 5100 | 17,277 | 0,058 |
| 7. | N | 14,01 | 63,29 | 77,4 | 0,209 | 1,365 |
| 8. | O | 15,999 | 54,8 | 90,188 | 0,458 | 1,141 |
| 9. | F | 18,998 | 53,53 | 85,01 | 0,343 | 1,240 |
| 10. | Ne | 20,179 | 24,48 | 27,1 | 0,027 | 1,544 |
| 11. | Na | 22,9897 | 370,96 | 1156,1 | 7,074 | 0,140 |
| 12. | Mg | 24,305 | 922 | 1363 | 3,758 | 0,260 |
| 13. | Al | 26,9815 | 933,52 | 2792,15 | 14,269 | 0,070 |
| 14. | Si | 28,086 | 1683 | 3538,15 | 13,682 | 0,073 |
| 15. | P | 30,973 | 317,3 | 553 | 1,576 | 0,565 |
| 16. | S | 32,064 | 388,36 | 717,82 | 2,128 | 0,439 |
| 17. | Cl | 35,453 | 172,17 | 239,18 | 0,392 | 1,198 |
| 18. | Ar | 39,948 | 83,78 | 87,29 | 0,018 | 1,553 |
| 19. | K | 39,102 | 336,8 | 1033,05 | 3,688 | 0,265 |
| 20. | Ca | 40,08 | 1112 | 1757 | 3,333 | 0,291 |
| 21. | Sc | 44,958 | 1814 | 3104 | 5,943 | 0,167 |
| 22. | Ti | 47,9 | 1933 | 3560 | 7,036 | 0,141 |
| 23. | V | 50,942 | 2160 | 3650 | 6,058 | 0,164 |
| 24. | Cr | 51,996 | 2130 | 2945 | 3,247 | 0,299 |
| 25. | Mn | 54,938 | 1517 | 2334,15 | 3,081 | 0,314 |
| 26. | Fe | 55,847 | 1808 | 3134,15 | 4,919 | 0,201 |
| 27. | Co | 58,9332 | 1768 | 3200,15 | 5,034 | 0,196 |
| 28. | Ni | 58,71 | 1726 | 3186,35 | 5,152 | 0,192 |
| 29. | Cu | 63,546 | 1356,95 | 2840 | 4,834 | 0,204 |
| 30. | Zn | 65,37 | 692,73 | 1180 | 1,544 | 0,575 |

*Table 1. (ending). Chemical elements*

*Values of estimation St and F are given at $T_{AT} = 273.15 K$*

| № | Element | Mass [g/mol] | $T_{MP}$ [K] | $T_{BP}$ [K] | *St* [ferg/g] | *F* [rad] |
|---|---|---|---|---|---|---|
| 31. | Ga | 69,72 | 302,93 | 2477,45 | 6,460 | 0,154 |
| 32. | Ge | 72,59 | 1210,6 | 3103 | 5,400 | 0,183 |
| 33. | As | 74,9216 | 1090 | 889 | -0,556 | 2,078 |
| 34. | Se | 78,96 | 490 | 958,1 | 1,228 | 0,683 |
| 35. | Br | 79,904 | 265,9 | 331,93 | 0,171 | 1,401 |
| 36. | Kr | 83,8 | 116,6 | 120,85 | 0,011 | 1,560 |
| 37. | Rb | 85,47 | 312,2 | 961 | 1,572 | 0,566 |
| 38. | Sr | 87,62 | 1042 | 1657 | 1,454 | 0,603 |
| 39. | Y | 88,905 | 1795 | 3611 | 4,231 | 0,232 |
| 40. | Zr | 91,224 | 2125 | 4650 | 5,733 | 0,173 |
| 41. | Nb | 92,906 | 2741 | 5015 | 5,070 | 0,195 |
| 42. | Mo | 95,94 | 2890 | 4885 | 4,307 | 0,228 |
| 43. | Tc | 98,9 | 2445 | 5150 | 5,665 | 0,175 |
| 44. | Ru | 101,07 | 2607,75 | 4423,15 | 3,721 | 0,263 |
| 45. | Rh | 102,905 | 2239 | 3968,15 | 3,481 | 0,280 |
| 46. | Pd | 106,42 | 1825 | 3213,15 | 2,702 | 0,354 |
| 47. | Ag | 107,868 | 1235,08 | 2435,15 | 2,304 | 0,409 |
| 48. | Cd | 112,411 | 594,1 | 1038 | 0,818 | 0,885 |
| 49. | In | 114,82 | 429,32 | 2353 | 3,470 | 0,281 |
| 50. | Sn | 118,69 | 505,118 | 2875,38 | 4,137 | 0,237 |
| 51. | Sb | 121,75 | 903,89 | 1860,35 | 1,627 | 0,551 |
| 52. | Te | 127,6 | 722,7 | 1263 | 0,877 | 0,851 |
| 53. | I | 126,904 | 386,7 | 457,5 | 0,116 | 1,456 |
| 54. | Xe | 131,29 | 161,3 | 166,1 | 0,008 | 1,563 |
| 55. | Cs | 132,905 | 301,55 | 942,45 | 0,999 | 0,786 |
| 56. | Ba | 137,34 | 1002 | 2170,65 | 1,763 | 0,516 |
| 57. | La | 138,91 | 1192 | 3737,35 | 3,795 | 0,258 |
| 72. | Hf | 178,49 | 2503 | 5470 | 3,443 | 0,283 |
| 73. | Ta | 180,9479 | 3269 | 5698 | 2,781 | 0,345 |
| 74. | W | 183,85 | 3680 | 5930 | 2,535 | 0,376 |
| 75. | Re | 186,207 | 3453 | 5900 | 2,722 | 0,352 |
| 76. | Os | 190,2 | 3327 | 5300 | 2,149 | 0,436 |
| 77. | Ir | 192,22 | 2683 | 4403 | 1,853 | 0,495 |
| 78. | Pt | 195,08 | 2045 | 4100 | 2,182 | 0,430 |
| 79. | Au | 196,966 | 1337,58 | 3080 | 1,832 | 0,500 |
| 80. | Hg | 200,59 | 234,28 | 629,73 | 0,408 | 1,183 |
| 81. | Tl | 204,38 | 576,7 | 1730 | 1,169 | 0,708 |
| 82. | Pb | 207,2 | 600,95 | 2013 | 1,412 | 0,616 |
| 83. | Bi | 208,98 | 544,5 | 1833 | 1,277 | 0,664 |
| 84. | Po | 209 | 527 | 1235 | 0,702 | 0,959 |
| 85. | At | 210 | 575 | 610 | 0,035 | 1,536 |
| 86. | Rn | 222 | 202 | 211,4 | 0,009 | 1,562 |
| 87. | Er | 223 | 300 | 950 | 0,604 | 1,028 |
| 88. | Ra | 226,02 | 973 | 1413 | 0,403 | 1,188 |

Let's display in the illustrative form results of calculation stability chemical elements estimations with 1 on 57 positions (*Fig. 8*).

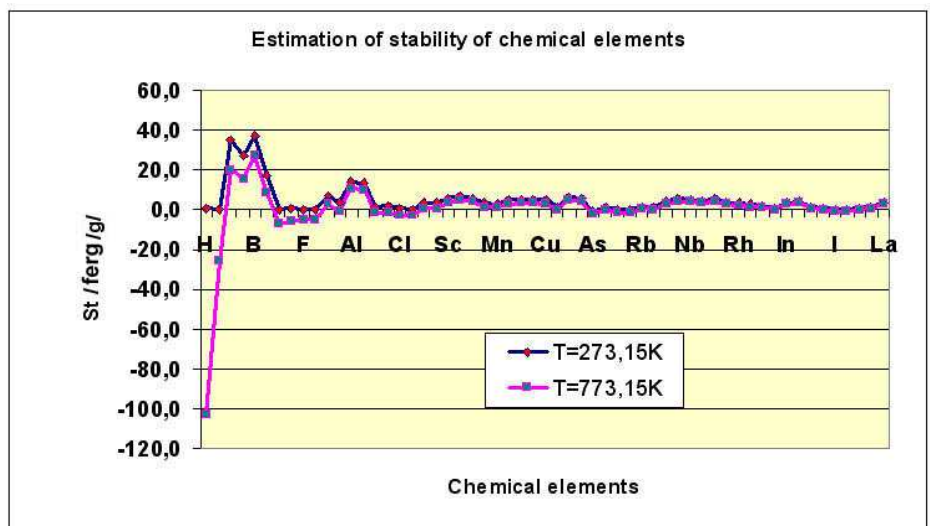

*Fig. 8. The comparative characteristic of chemical elements stability*

*With the purpose comparison St estimations are brought in an illustrative material received at $T_{AT} = 273.15K$ and $T_{AT} = 773.15K$. On a rating scale negative values St correspond to a low level of stability.*

In a considered number of chemical elements the arsenic (As) is allocated with physical properties. Stability of this chemical element is shown distinctly enough. The subgroup of arsenic is insignificant. Into this group enter: antimony and bismuth. In an earth's crust these substances contain in small quantities:

$$As - 1*10^{-4}\%$$
$$Sb - 5*10^{-6}\%$$
$$Bi - 2*10^{-6}\%$$

Arsenic on property of stability is allocated in this group. In a free state elements of a subgroup of arsenic have a metal appearance, well spend heat and electricity. In the nature these elements meet in the form of sulfide minerals. Impurities of all three elements often meet in ores of various metals. Arsenic is capable to exist in several allotropic forms, from which steadiest "grey" form. At fast cooling of arsenic vapors the «yellow arsenic» $As_4$ turns out. The stereochemical molecular structure of «yellow arsenic» has structure of a correct tetrahedron. On air such chemical compound easily is oxidized, as is reflected in property of stability. Settlement values of stability for *As* is:

$$St_{T=273.15K} = -0.556 [ferg/g]$$
$$St_{T=773.15K} = -1.938 [ferg/g]$$

The received estimations testify that with growth of external temperature, stability of an element decreases. It is marked and in practice of researches of substance. Change of structure for this substance can occur and in usual states, for example, under action of light. Under action of light the «yellow arsenic» passes in to the «grey arsenic», widely widespread in the nature. At sublimation As in hydrogen stream forms amorphous – «black arsenic» which is not oxidized on air, but at temperature $T_{AT} > 543.15K$ passes in to the «grey form».

About a chemical element arsenic allows to speak the general representations about natural weak stability of element external factors. Ability of arsenic to self-destruction will well be coordinated with the received estimations of stability.

Let's consider estimations fugacity for the chemical elements presented in periodic system with 1 on 57 positions (*Fig. 9*). Settlement estimations are presented for two values of external temperature: $T_{AT} = 273.15K$, $T_{AT} = 773.15K$. The first elements in this number, as well as follows, to expect, elements are presented: hydrogen and helium. Following elements - nitrogen, oxygen, fluorine and neon, are formed compact group. The analysis of the received results of calculation distinctly shows periodicity of estimations fugacity for the chosen chemical elements.

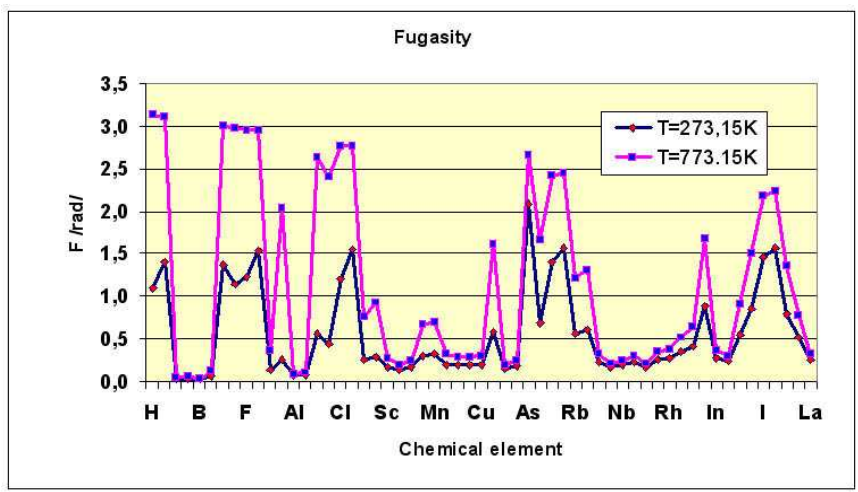

*Fig. 9. The comparative characteristic fugacity chemical elements*

Let's expand representations about fugacity and we shall consider a gaseous mix of different elements. Air has complex structure. In air allocate three basic components: oxygen (21 %), nitrogen (78 %) and inert gaseous (1 %). Is of interest to specify the received estimations on well studied objects – inert gaseous: He, Ne, Ar, Kr, Xe, and Ra. We shall allocate these objects in separate group and we shall present corresponding estimations in *Table 2.*

*Table 2. Rare gases*

| № | Element | $T_{AT}$=223,15K | | $T_{AT}$=273,15K | | $T_{AT}$=773,15K | |
|---|---|---|---|---|---|---|---|
| | | St [ferg/g] | F [rad] | St [ferg/g] | F [rad] | St [ferg/g] | F [rad] |
| 1 | H | 10,868 | 0,092 | 0,512 | 1,098 | -103,056 | 3,132 |
| 2 | He | 2,757 | 0,348 | 0,169 | 1,403 | -25,710 | 3,103 |
| 18 | Ar | 0,277 | 1,300 | 0,018 | 1,553 | -2,574 | 2,771 |
| 36 | Kr | 0,134 | 1,437 | 0,011 | 1,560 | -1,225 | 2,457 |
| 54 | Xe | 0,086 | 1,485 | 0,008 | 1,563 | -0,781 | 2,234 |
| 86 | Rn | 0,055 | 1,515 | 0,009 | 1,562 | -0,458 | 2,000 |

The received estimations we shall present in the illustrative form on *Fig. 10*. We shall note that the first place borrows hydrogen, and last place belongs to radon. We shall execute similar calculations for definition of estimation fugacity rare gases. We shall illustrate results on *Fig. 10*. In a number of inert gaseous special positions borrow helium. Among inert gaseous, helium can be observed in a solid state, but at pressure nearly $P = 25 atm$. This feature of helium is shown in the calculated estimation fugacity. In calculation fugacity for helium at different values of ambient temperature, this gaseous is on the second place after hydrogen.

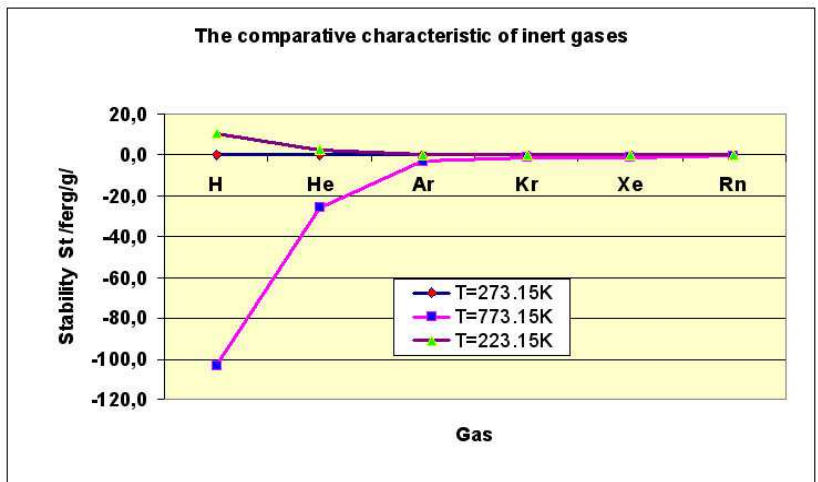

*Fig.10. Distribution of stability estimations in a number of inert gaseous*

With rise in ambient temperature fugacity helium grows more intensively, than at other gases. Exception is hydrogen, for which estimation fugacity always above, than at helium. In the general portrait of distribution of estimations fugacity consideration of a trend is of interest at low temperature.

For example, at $T_{AT} = 223.15K$ increase of estimation St and downturn of estimation F for hydrogen and helium is distinctly observed.

At the selected temperature ($T_{AT} = 223.15K$) for other inert gaseous the level of an estimation fugacity goes down, but is insignificant. So, for radon downturn of temperature changes an estimation fugacity which remains practically constant a little. On the contrary, for this gas rise in ambient temperature essentially raises a level of an estimation fugacity (*Fig. 11*).

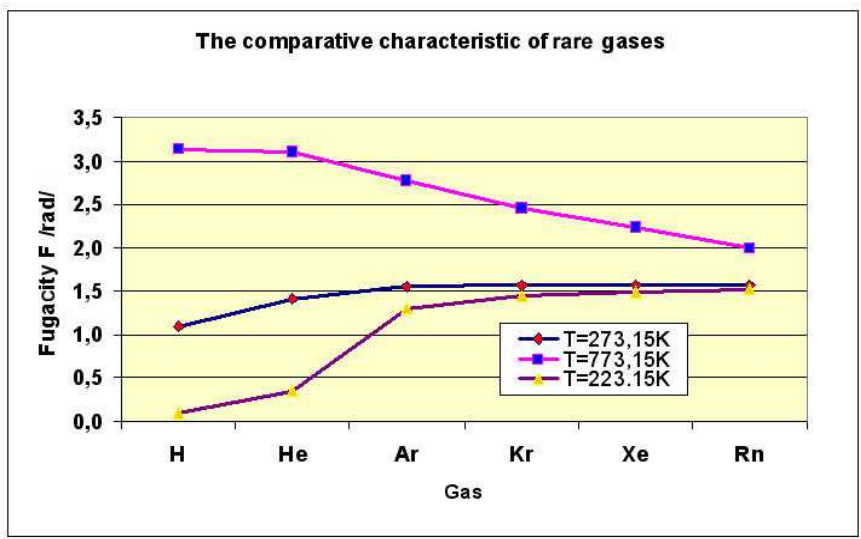

*Fig. 11. Distribution of estimations fugacity in a number of inert gaseous*

Having received the general representations about distribution of estimations fugacity in a number of chemical elements, we shall consider the ordered number of estimations fugacity. On the one hand construction such of some will clear a position of a chemical element in a number fugacity, and on the other hand, will allow establishing the elements possessing a high parameter fugacity. Thus it is possible to note, that construction of some estimations fugacity opens ample opportunities on studying mutual communication with other typical properties of the chemical elements widely presented in various researches.

Analyzing the received estimations fugacity it is possible to state a hypothesis that at change of external temperature change of a position of elements will be observed. However, if to consider rather small fluctuations of external temperature for some chemical elements change of number of a position will be small. On the contrary, for the chemical elements belonging to a category of gases, change of number of a position will be greater. We shall any way choose a range of change of external temperature and we shall lead calculation of estimations fugacity for chemical elements. Results of calculation we shall present in *Table 3.*

*Table 3. Phenomenological sequence estimations fugacity chemical elements*

| № | $T_{AT}=223.15K$ | | $T_{AT}=273.15K$ | | $T_{AT}=293.15K$ | | $T_{AT}=773.15K$ | |
|---|---|---|---|---|---|---|---|---|
| | F [rad] | Element | F [rad] | Element | F [rad] | Element | F [rad] | Element |
| **1.** | 1,966 | As | 2,078 | As | 2,873 | H | 3,132 | H |
| **2.** | 1,515 | Rn | 1,563 | Xe | 2,285 | He | 3,103 | He |
| **3.** | 1,487 | At | 1,562 | Rn | 2,119 | As | 3,003 | N |
| **4.** | 1,485 | Xe | 1,560 | Kr | 1,747 | Ne | 2,977 | O |
| **5.** | 1,437 | Kr | 1,553 | Ar | 1,658 | N | 2,948 | F |
| **6.** | 1,376 | I | 1,544 | Ne | 1,656 | Ar | 2,948 | Ne |

*Table 3 (continuation). Phenomenological sequence estimations fugacity chemical elements*

| №   | $T_{AT}=223.15K$ | | $T_{AT}=273.15K$ | | $T_{AT}=293.15K$ | | $T_{AT}=773.15K$ | |
|     | F [rad] | Element | F [rad] | Element | F [rad] | Element | F [rad] | Element |
| --- | --- | --- | --- | --- | --- | --- | --- | --- |
| 7.  | 1,300 | Ar | 1,536 | At | 1,610 | Kr | 2,771 | Ar |
| 8.  | 1,279 | Br | 1,456 | I  | 1,595 | Xe | 2,765 | Cl |
| 9.  | 1,149 | Ra | 1,403 | He | 1,581 | Rn | 2,665 | As |
| 10. | 1,140 | Hg | 1,401 | Br | 1,556 | At | 2,627 | P  |
| 11. | 1,076 | Ne | 1,365 | N  | 1,488 | I  | 2,457 | Kr |
| 12. | 0,994 | Fr | 1,240 | F  | 1,452 | Br | 2,415 | Br |
| 13. | 0,971 | Cl | 1,198 | Cl | 1,446 | F  | 2,405 | S  |
| 14. | 0,927 | Po | 1,188 | Ra | 1,374 | O  | 2,234 | Xe |
| 15. | 0,844 | F  | 1,183 | Hg | 1,303 | Cl | 2,182 | I  |
| 16. | 0,832 | Cd | 1,141 | O  | 1,203 | Ra | 2,037 | Mg |
| 17. | 0,812 | N  | 1,098 | H  | 1,201 | Hg | 2,001 | At |
| 18. | 0,807 | Te | 1,028 | Fr | 1,041 | Fr | 2,000 | Rn |
| 19. | 0,748 | Cs | 0,959 | Po | 0,972 | Po | 1,678 | Hg |
| 20. | 0,735 | O  | 0,885 | Cd | 0,908 | Cd | 1,674 | Cd |
| 21. | 0,687 | Tl | 0,851 | Te | 0,869 | Te | 1,654 | Se |
| 22. | 0,646 | Bi | 0,786 | Cs | 0,802 | Cs | 1,626 | Ra |
| 23. | 0,634 | Se | 0,708 | Tl | 0,716 | Tl | 1,611 | Zn |
| 24. | 0,600 | Pb | 0,683 | Se | 0,705 | Se | 1,505 | Te |
| 25. | 0,567 | Sr | 0,664 | Bi | 0,672 | Bi | 1,432 | Fr |
| 26. | 0,533 | Rb | 0,616 | Pb | 0,623 | Pb | 1,368 | Po |
| 27. | 0,531 | Zn | 0,603 | Sr | 0,618 | Sr | 1,355 | Cs |
| 28. | 0,529 | Sb | 0,575 | Zn | 0,606 | P  | 1,305 | Sr |
| 29. | 0,498 | Ba | 0,566 | Rb | 0,594 | Zn | 1,225 | Rb |
| 30. | 0,488 | Au | 0,565 | P  | 0,581 | Rb | 0,986 | Tl |
| 31. | 0,483 | Ir | 0,551 | Sb | 0,561 | Sb | 0,928 | Ca |
| 32. | 0,482 | P  | 0,516 | Ba | 0,524 | Ba | 0,910 | Sb |
| 33. | 0,426 | Os | 0,500 | Au | 0,504 | Au | 0,907 | Bi |
| 34. | 0,421 | Pt | 0,495 | Ir | 0,500 | Ir | 0,832 | Pb |
| 35. | 0,395 | Ag | 0,439 | S  | 0,464 | S  | 0,781 | Ba |
| 36. | 0,387 | S  | 0,436 | Os | 0,439 | Os | 0,766 | K  |
| 37. | 0,368 | W  | 0,430 | Pt | 0,433 | Pt | 0,696 | Mn |
| 38. | 0,348 | He | 0,409 | Ag | 0,416 | Ag | 0,673 | Cr |
| 39. | 0,346 | Re | 0,376 | W  | 0,379 | W  | 0,653 | Au |
| 40. | 0,343 | Pd | 0,354 | Pd | 0,359 | Pd | 0,650 | Ir |
| 41. | 0,339 | Ta | 0,352 | Re | 0,355 | Re | 0,640 | Ag |
| 42. | 0,297 | Mn | 0,345 | Ta | 0,348 | Ta | 0,557 | Os |
| 43. | 0,282 | Cr | 0,314 | Mn | 0,321 | Mn | 0,545 | Pt |
| 44. | 0,278 | Hf | 0,299 | Cr | 0,306 | Cr | 0,524 | Pd |
| 45. | 0,274 | In | 0,291 | Ca | 0,300 | Ca | 0,469 | W  |

*Table 3 (ending). Phenomenological sequence estimations fugacity chemical elements*

| № | $T_{AT}=223.15K$ | | $T_{AT}=273.15K$ | | $T_{AT}=293.15K$ | | $T_{AT}=773.15K$ | |
| --- | --- | --- | --- | --- | --- | --- | --- | --- |
| | F [rad] | Element | F [rad] | Element | F [rad] | Element | F [rad] | Element |
| 46. | 0,272 | Rh | 0,283 | Hf | 0,284 | Hf | 0,433 | Re |
| 47. | 0,272 | Ca | 0,281 | In | 0,283 | In | 0,425 | Ta |
| 48. | 0,256 | Ru | 0,280 | Rh | 0,283 | Rh | 0,384 | Rh |
| 49. | 0,253 | La | 0,265 | K | 0,272 | K | 0,371 | In |
| 50. | 0,248 | K | 0,263 | Ru | 0,272 | Mg | 0,371 | Na |
| 51. | 0,235 | Mg | 0,260 | Mg | 0,265 | Ru | 0,355 | Ru |
| 52. | 0,232 | Sn | 0,258 | La | 0,260 | La | 0,336 | Hf |
| 53. | 0,226 | Y | 0,237 | Sn | 0,239 | Sn | 0,317 | La |
| 54. | 0,223 | Mo | 0,232 | Y | 0,235 | Y | 0,315 | Fe |
| 55. | 0,198 | Cu | 0,228 | Mo | 0,230 | Mo | 0,315 | Y |
| 56. | 0,193 | Fe | 0,204 | Cu | 0,207 | Cu | 0,302 | Cu |
| 57. | 0,191 | Nb | 0,201 | Fe | 0,204 | Fe | 0,300 | Mo |
| 58. | 0,190 | Co | 0,196 | Co | 0,199 | Co | 0,297 | Sn |
| 59. | 0,186 | Ni | 0,195 | Nb | 0,196 | Nb | 0,296 | Co |
| 60. | 0,178 | Ge | 0,192 | Ni | 0,194 | Ni | 0,287 | Ni |
| 61. | 0,172 | Tc | 0,183 | Ge | 0,185 | Ge | 0,268 | Sc |
| 62. | 0,169 | Zr | 0,175 | Tc | 0,176 | Tc | 0,248 | Nb |
| 63. | 0,161 | Sc | 0,173 | Zr | 0,174 | Zr | 0,247 | Ge |
| 64. | 0,158 | V | 0,167 | Sc | 0,169 | Sc | 0,243 | V |
| 65. | 0,150 | Ga | 0,164 | V | 0,166 | V | 0,214 | Zr |
| 66. | 0,137 | Ti | 0,154 | Ga | 0,155 | Ga | 0,213 | Tc |
| 67. | 0,132 | Na | 0,141 | Ti | 0,144 | Na | 0,202 | Ti |
| 68. | 0,092 | H | 0,140 | Na | 0,143 | Ti | 0,198 | Ga |
| 69. | 0,071 | Si | 0,073 | Si | 0,074 | Si | 0,115 | C |
| 70. | 0,068 | Al | 0,070 | Al | 0,071 | Al | 0,100 | Si |
| 71. | 0,055 | C | 0,058 | C | 0,059 | C | 0,096 | Al |
| 72. | 0,035 | Be | 0,037 | Be | 0,037 | Be | 0,063 | Be |
| 73. | 0,028 | Li | 0,029 | Li | 0,029 | Li | 0,050 | Li |
| 74. | 0,026 | B | 0,027 | B | 0,027 | B | 0,037 | B |

The result of calculations of estimations fugacity is presented by a series of figures (***Fig.12 - Fig.15***).

The stated hypothesis about an opportunity of detection of some stable chemical elements possessing practically constant and small values of estimations fugacity on the selected range of temperatures has proved to be true. The first chemical element can be named carbon, which natural formation well-known – diamond. Lower estimations fugacities are established for chemical elements: Be, Li, B.

In the pure state Be represents easy, solid, proof to corrosion, grey-steel color metal. Chemical element Li is present at minerals which representatives are petalite and spodumene. Typical representative Li of mineral origin is mica. Mica — is one of the most widespread minerals of rocks. This mineral possesses a high degree of stability. Chemical element B is known in the form of amorphous green - brown powder or the fine crystals having metal shine and on hardness almost not conceding diamond.

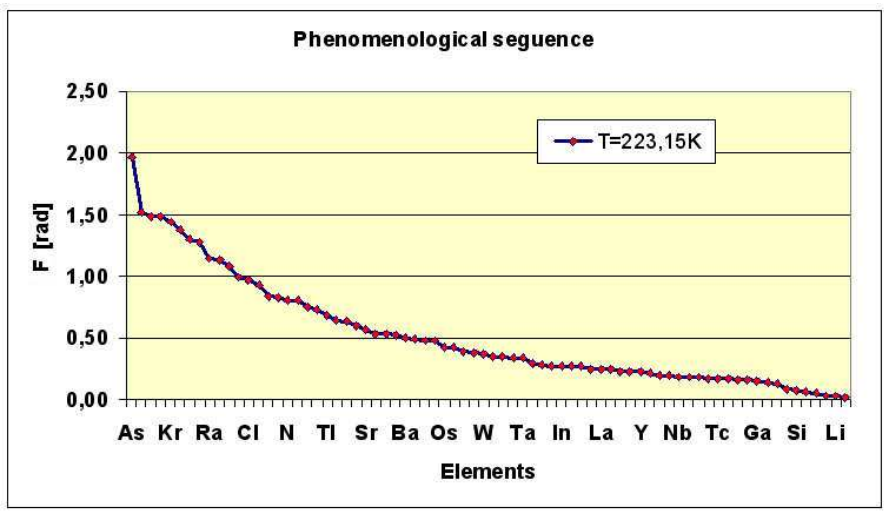

***Fig. 12.*** *Estimation chemical elements fugacity ($T_{AT}$=223.15K)*

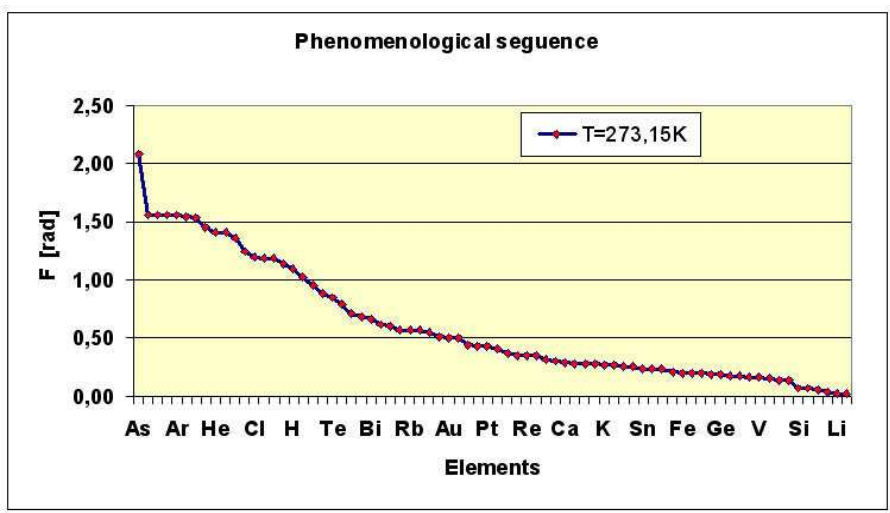

***Fig. 13.*** *Estimation chemical elements fugacity ($T_{AT}$=273.15K)*

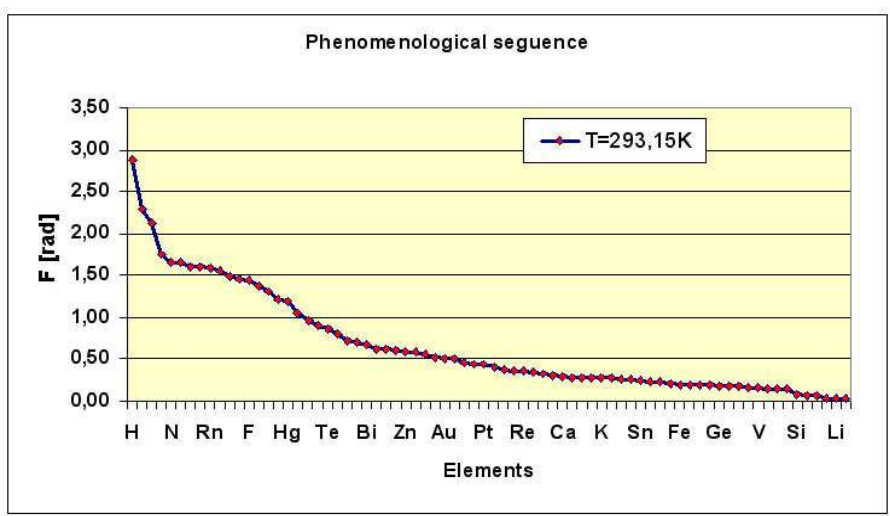

***Fig. 14.*** *Estimation chemical elements fugacity ($T_{AT}$=293.15K)*

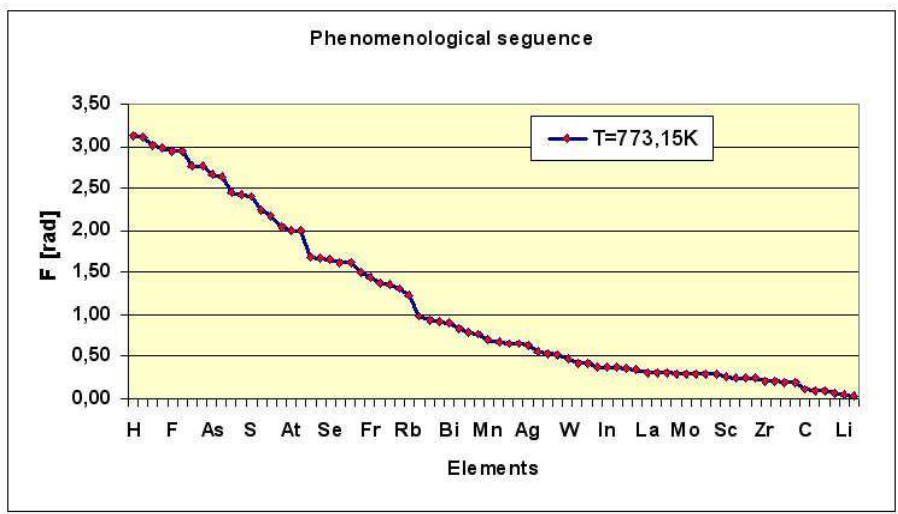

***Fig. 15.*** *Estimation chemical elements fugacity ($T_{AT}$=773.15K)*

## 8. RESULTS AND DISCUSSION

Let's lead comparison of the taught results with classical data of calculation vaporization enthalpy ΔH. We shall consider estimations enthalpy and stability St for chemical elements in ***Table 4*** [1].

***Table 4.*** *Estimations series of ΔH and St ($T_{AT}$=273.15K)*

| № | Element | Vaporization enthalpy H [kJ/mol] | St $60*10^{-22}$ [J/g] | № | Element | Vaporization enthalpy H [kJ/mol] | St $60*10^{-22}$ [J/g] |
|---|---|---|---|---|---|---|---|
| 1 | H | 0,46 | 30,7 | 20 | Ca | 150,6 | 200,0 |
| 2 | He | 0,082 | 10,1 | 21 | Sc | 376,1 | 356,6 |
| 3 | Li | 147,7 | 2088,3 | 22 | Ti | 425,5 | 422,1 |
| 4 | Be | 308,8 | 1634,2 | 23 | V | 459,7 | 363,5 |
| 5 | B | 504,5 | 2208,3 | 24 | Cr | 341,8 | 194,8 |
| 6 | C | 710,9 | 1036,6 | 25 | Mn | 220,5 | 184,9 |
| 7 | N | 5,58 | 12,5 | 26 | Fe | 340,2 | 295,1 |
| 8 | O | 6,82 | 27,5 | 27 | Co | 382,4 | 302,0 |
| 9 | F | 3,26 | 20,6 | 28 | Ni | 374,8 | 309,1 |
| 10 | Ne | 1,736 | 1,6 | 29 | Cu | 306,7 | 290,0 |
| 11 | Na | 99,2 | 424,4 | 30 | Zn | 114,2 | 92,6 |
| 12 | Mg | 127,6 | 225,5 | 31 | Ga | 270,3 | 387,6 |
| 13 | Al | 290,8 | 856,1 | 32 | Ge | 327,6 | 324,0 |
| 14 | Si | 383,3 | 820,9 | 33 | As | 31,9 | -33,3 |
| 15 | P | 51,9 | 94,6 | 34 | Se | 90 | 73,7 |
| 16 | S | 9,62 | 127,7 | 35 | Br | 30,5 | 10,3 |
| 17 | Cl | 20,42 | 23,5 | 36 | Kr | 9,05 | 0,6 |
| 18 | Ar | 6,53 | 1,1 | 37 | Rb | 75,7 | 94,3 |
| 19 | K | 79,1 | 221,3 | 38 | Sr | 154,4 | 87,2 |

***Table 4 (ending).** Estimations series of ΔH and St ($T_{AT}$=273.15K)*

| № | Element | Vaporization enthalpy H [kJ/mol] | St $60*10^{-22}$ [J/g] | № | Element | Vaporization enthalpy H [kJ/mol] | St $60*10^{-22}$ [J/g] |
|---|---|---|---|---|---|---|---|
| 39 | Y | 367,4 | 253,9 | 57 | La | 402,1 | 227,7 |
| 40 | Zr | 566,7 | 344,0 | 72 | Hf | 570,7 | 206,6 |
| 41 | Nb | 680,19 | 304,2 | 73 | Ta | 758,22 | 166,8 |
| 42 | Mo | 589,9 | 258,4 | 74 | W | 824,2 | 152,1 |
| 43 | Tc | 585,2 | 339,9 | 75 | Re | 704,25 | 163,3 |
| 44 | Ru | 567 | 223,2 | 76 | Os | 738,06 | 128,9 |
| 45 | Rh | 494,3 | 208,8 | 77 | Ir | 612,1 | 111,2 |
| 46 | Pd | 361,5 | 162,1 | 78 | Pt | 469 | 130,9 |
| 47 | Ag | 257,7 | 138,3 | 79 | Au | 343,1 | 109,9 |
| 48 | Cd | 100,0 | 49,1 | 80 | Hg | 59,11 | 24,5 |
| 49 | In | 231,8 | 208,2 | 81 | Tl | 166,1 | 70,1 |
| 50 | Sn | 296,2 | 248,2 | 82 | Pb | 177,8 | 84,7 |
| 51 | Sb | 165,8 | 97,6 | 83 | Bi | 179,1 | 76,6 |
| 52 | Te | 104,6 | 52,6 | 84 | Po | 100,8 | 42,1 |
| 53 | I | 41,67 | 6,9 | 85 | At | - | 2,1 |
| 54 | Xe | 12,65 | 0,5 | 86 | Rn | 18,1 | 0,5 |
| 55 | Cs | 66,5 | 59,9 | 87 | Fr | - | 36,2 |
| 56 | Ba | 150,9 | 105,8 | 88 | Ra | 136,7 | 24,2 |

Let's note that values ΔH correspond to data of the physical directory. We shall lead comparison of two thermodynamic estimations in the illustrative form (***Fig. 16***).

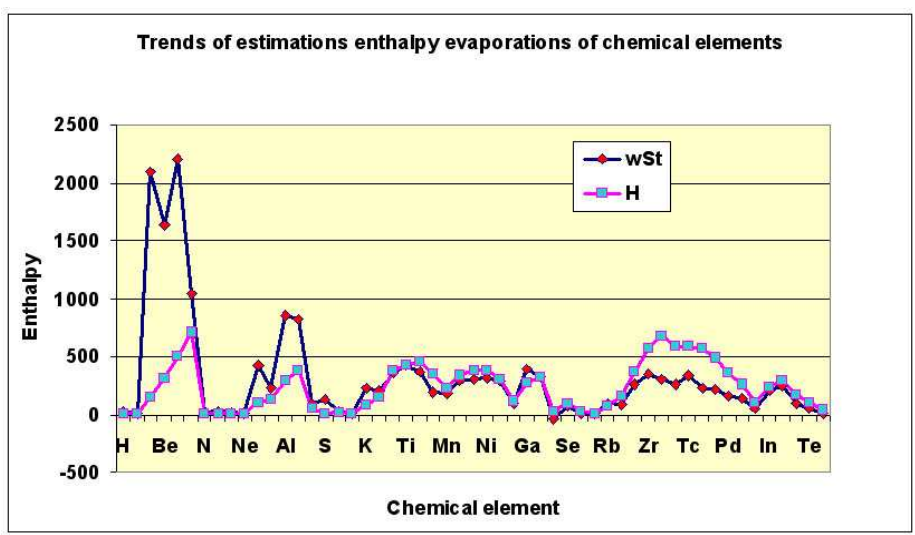

***Fig. 16.** Comparison of settlement and theoretical estimations where: wSt = $60*10^{-22}$ [J/mol]*

Analyzing schedules, we establish similarity of estimations ΔH and St. The received result allows noting efficiency of used physical process model.

Now questions of natural disintegration of substance, and it are formally possible so to characterize fugacity, are considered in different areas of researches. Using known physical representations about this process, various models which research allows establishing estimations fugacity are created. In the most widespread representation usually use classical equation D. Mendeleyev for a state of ideal gaseous. The positive beginning here is that the gaseous aggregative state is allocated. The gaseous fraction of a state of object, within the limits of numerous models, is considered as a primary basis for calculation of estimations fugacity. As at formation of initial model structural features of construction of objects are not considered, estimations fugacity aspire to present in a dimensionless kind, using for this purpose normalized scale of values [0; 1]. From the point of view of practice such approach allows establishing roughly estimations fugacity for of some substances and can be considered as outline. In addition to it is possible to tell that such calculations are complex enough and possess small illustrative at transition from one class of substances to another.

Considering, that fugacity – process connected with expressiveness of external factors and, first of all, with temperature, logically correctly to discuss a question on stability of the object possessing in certain parameters of a state. In that case it is possible to speak about molecular weight, density, a thermal capacity and a stereochemical structure of object. Identifying some object with the certain physical parameters of a state, there is an opportunity to investigate object - to establish its some properties on model.

Constructing of the model that reflecting events interesting us, which can hypothetically occur to object, is accompanied by search of analogues of real processes. From these positions the concept entered by the last century fugacity has set of illustrative and real examples in practice. Saved up since then, practical and theoretical, experience allows considering the given process from a position of thermodynamics within the limits of which the accent is put on initial representations about interaction of any object with a medium. Following these principles it is possible to establish, that time of existence – time of a life of any material object certainly and is defined as an internal structure (internal energy) object, and factors of medium. Allocating as the significant factor – ambient temperature which naturally influences object, it is possible to create model of stability of object. It is obvious, that stability of internal energy of object in system of interaction with a medium is not possible. In that case loss of internal energy by object corresponds to disintegration. Disintegration of object is characterized by loss of weight which is injected in a medium, by means of work of the external forces made above object. It is obvious, that the greater work is made above object, the greater expressiveness of process of disintegration.

From a position of physics of the given process it is possible to tell, that the formalism of the description is limited by two extreme concepts: temperature value of absolute zero and over high values of ambient temperature. It is obvious, that in the first case pertinently to speak about absence of display of the external factor – temperatures, and in the second, on the contrary, about the big influence on object of external temperature. According to it there is a formation of a scale fugacity. Allocating the second case it is possible to find easily set of examples of the ionized state of various substances – the plasma characterized by great value of temperature. The ionized state of substance is identified with a heat of medium that is why consideration of process fugacity gets the generalized character practically for any object.

The approach considered in given work to formation of an estimation fugacity various objects is based on positions of thermodynamics. Using the general representations about thermodynamic function of a state, the model of development of disintegration of the object possessing certain properties is formed. Bringing to a focus to unique properties of object, physical parameters of a state which assume as basis models of succession of events in system "object – medium" are allocated. Using modeling representations about succession of events in system, calculations of estimations fugacity for typical chemical elements and inert gaseous are spent. Owing to the good level of scrutiny of this theme of researches which are not demanding additional explanatory and references to a plenty of practical works, the received results of calculation of estimations fugacity allow to generate a phenomenological number for chemical elements and inert gaseous.